\documentclass[prl,twocolumn,letterpaper,showpacs,groupedaddress,superscriptaddress,nofootinbib,floatfix,preprintnumbers,tightenlines]{revtex4-1}
\usepackage[utf8]{inputenc}
\usepackage{bm}
\usepackage{hyperref}
\usepackage{amssymb}
\usepackage{amsmath}
\usepackage{epsfig}
\usepackage{slashed}
\usepackage{comment}
\usepackage{mathtools}
\usepackage{datetime}
\usepackage{rotating}
\usepackage{relsize}

\usepackage{stackengine}
\stackMath
\usepackage[english]{babel}
\usepackage[utf8]{inputenc}
\usepackage{fancyhdr}
\newcommand{\half}{\frac{1}{2}}
\newcommand{\be}{\begin{equation}}
\newcommand{\ee}{\end{equation}}
\newcommand\beq{\begin{eqnarray}}
\newcommand\eeq{\end{eqnarray}} 
\newcommand\eqn[1]{\label{eq:#1}} 
\newcommand\eq[1]{eq.~(\ref{eq:#1})}

\newcommand{\CJ}{{\cal J}}

\newcommand{\CQ}{{\cal Q}}

\newcommand{\CL}{{\cal L}}

\newcommand{\Tr}{{\rm Tr\,}}

\newcommand\ket[1]{\,| #1 \rangle}

\newcommand{\mybar}[1]%
        {\kern 0.6pt\overline{\kern -0.6pt#1\kern -0.6pt}\kern 0.6pt}

\makeatletter
\renewcommand*\env@matrix[1][\arraystretch]{%
  \edef\arraystretch{#1}%
  \hskip -\arraycolsep
  \let\@ifnextchar\new@ifnextchar
  \array{*\c@MaxMatrixCols c}}
\makeatother
\begin{document}

\title{Fractional quantum Hall effect in a relativistic field theory}

\preprint{INT-PUB-19-054}

\author{David B. Kaplan}
 \email{dbkaplan@uw.edu}
\affiliation{Institute for Nuclear Theory, Box 351550, University of Washington, Seattle, WA 98195-1550}

\author{Srimoyee Sen}
 \email{srimoyee08@gmail.com}
\affiliation{Department of Physics, Iowa State University, Ames, IA 50011-3160}

\begin{abstract}
   We construct a class of 2+1 dimensional relativistic quantum field theories which exhibit the Fractional Quantum Hall Effect in the infrared, both in the continuum and on the lattice. The UV completion consists of a perturbative $U(1)\times U(1) $ gauge theory with only integer-charged fields, while the low energy spectrum consists of  nontrivial topological phases supporting fractional currents, fractionally charged chiral surface modes and bulk anyonic excitations.  Exotic phenomena such as a Fractional Quantum Spin Hall Effect can arise in such models.

\end{abstract}

\maketitle

\section{Introduction}

Topological materials have been discussed in the context of lattice quantum field theories since the early 1990s.  In particular, the theory of  domain wall fermions (DWF) \cite{Kaplan:1992bt} -- designed to  optimally realize  chiral symmetry on a lattice within the constraints of the Nielsen-Ninomiya theorem \cite{Nielsen:1980rz} -- is an example of a topological insulator, a gapped  symmetry protected topological phase of matter \cite{wen1989vacuum}. In the  DWF theory, Chern-Simons (CS) currents in the gapped bulk play the role of quantized Hall currents  \cite{Callan:1984sa}, while the chiral fermions bound to the surfaces  \cite{Jackiw:1975fn}  are equivalent  to the edge states found in the integer quantum Hall system, albeit generalized to non-abelian gauge theories. A field theoretic generalization of the TKNN calculation \cite{thouless1982quantized} was given in  Ref.~\cite{Golterman:1992ub}  where it was shown that the quantization of the CS operator coefficient (responsible for the quantization of the bulk current)  had a topological origin, and that transitions between these topological phases were responsible for abrupt changes seen in the spectrum of chiral edge states \cite{Jansen:1992tw}.
In particular, the Feynman diagram calculation of the CS coefficient  in $d+1$ spacetime dimensions was shown to compute the winding number of the  map from momentum space to the sphere $S^{d+1}$, explaining its quantization and insensitivity to  parameters in the Lagrangian.  Furthermore, the constructions of anomaly-free representations for  chiral edge states given in \cite{Kaplan:1992bt,Jansen:1992yj} (for the purpose of regulating chiral gauge theories)  have vanishing gauged but nonzero quantized  chiral flavor currents in the bulk, and are realizations of the quantum spin Hall effect\footnote{The simplest anomaly-free theory consists of a Dirac fermion at each edge, as in the examples of the SQHE seen experimentally \cite{konig2007quantum,hsieh2008topological}, although less trivial examples of anomaly cancellation have been studied numerically \cite{Jansen:1992yj}.}. 
 In yet another correspondence between relativistic and condensed matter topological materials, Majorana fermion edge states were introduced  in the relativistic context  \cite{Kaplan:1999jn}  and used for the numerical simulation of gauginos in  supersymmetric gauge theory \cite{Endres:2009yp}, analogous to the Majorana modes of condensed matter systems \cite{kitaev2001unpaired}.

Despite this long list of topological phases found in condensed matter that also play a role in relativistic quantum field theory, notably missing to date is the fractional quantum Hall effect (FQHE).    In this Letter we provide an example of such a theory,   by which we mean a UV-complete $2+1$ dimensional, Lorentz-invariant gauge theory of  integer-charged matter fields, where the low energy theory at zero chemical potential is characterized by topological phases with fractionally charged chiral edge modes and a fractional coefficient for the Chern-Simons operator in the bulk.  The framework is perturbative, and provides examples of other exotic phenomena, such as a fractional quantum spin Hall effect.  We show explicitly the emergence of fractionally charged chiral edge modes as well as bulk excitations with fractional charge and statistics.  Our construction is motivated by existing effective field theory descriptions of IR phenomena in condensed matter   systems (see, for example, Ref~\cite{Witten:2015aoa, Tong:2016kpv}).

Our theory is a  $U(1) \times U(1)$ gauge theory in $2+1$ dimensions, with gauge bosons $A_\alpha$ (the ``photon") and $ Z_\alpha$, and three types of fermions $\psi$, $\chi$ and $\omega$ with charge assignments as shown in Table~\ref{tab:charges}.  For simplicity we take  all fermion masses to have the same magnitude, $|m_i|= M$, where $M$ is some positive mass scale.   
Each fermion is assigned an integer  flavor number $n_\psi$, $n_\chi$ and $n_\omega$ respectively, where $|n_i|$ denotes the number of degenerate flavors   of fermion type $i$, while the sign of $n_i$ equals the sign of the fermion mass, $n_i/|n_i| \equiv m_i/|m_i|$.  The Lagrangian then consists of the usual Dirac  terms for the fermions with covariant derivative $D_\alpha = \partial_\alpha + i q_A A_\alpha + i q_Z Z_\alpha$, as well as Maxwell terms for the gauge bosons, 
$-\frac{1}{4e^2} F_{\alpha\beta}F^{\alpha\beta}$, $ -\frac{1}{4g^2} Z_{\alpha\beta}Z^{\alpha\beta}$ where $Z_{\alpha\beta}$ is the field strength for the $Z_\alpha$ gauge boson and $g$ is its coupling constant.  The $q_A$ and $q_Z$ charges we choose for the fermions are given in Table~\ref{tab:charges}.

 \begin{table}[t]
\caption{\label{tab:charges} Dirac fermions labeled by $U(1)_A\times U(1)_Z$  charge assignments $\{q_A,\,q_Z\}$. }
\begin{ruledtabular}
\begin{tabular}{lccc}
&$n_\text{flavor}$ &$q_A$&$q_Z $\\
\hline
$\psi$ &$n_\psi$ & 1 & 0 \\
$\chi$ &$n_\chi$ & 0 & 1 \\
$\omega$ &$n_\omega$ & 1 & 1 \\
\end{tabular}
\end{ruledtabular}
\end{table}

 The topological phase structure  of the theory depends on momentum space being compact, and so we consider two different regularizations in this Letter with quite different topological phases: we first employ Pauli-Villars fields, and then later consider a lattice regularization with Wilson fermions, the former compactifying momentum space to a sphere and the latter to a torus.   There is one Pauli-Villars field of mass $\Lambda$ for each fermion in the theory with mass $M$, with the same charge assignment but opposite statistics and opposite sign mass, $\Lambda/|\Lambda| = - M/|M|$.  The choice of having a relative sign between the mass of each fermion and its regulator allows the theory to be in  a nontrivial topological phase, as we will discuss below.
  With our normalization of the gauge fields, and with $\hbar=c=1$,  the theory  has four different mass scales: $ \Lambda$, $M$, $g^2$ and $e^2$, which we choose to obey the following inequalities, 
 \beq
 |\Lambda| \gg |M| \gg  g^2 \gg e^2\ ,
\eqn{hierarchy} \eeq
where the hierarchies are considered to be much bigger than the numbers of flavors, $|n_{\psi,\chi,\omega}|$.
   
We now construct the effective low energy theory for this system. Given our hierarchy of scales in \eq{hierarchy}, we  first integrate out the massive fermions and their regulator fields.  The resultant theory is gapped, with the exception of massless chiral edge states if the system has a boundary.   The next heaviest state is the $Z$ boson, and so we next integrate out that field, yielding the effective theory for just the photon and the edge states.  
 
First we consider the theory without a boundary, postponing the discussion of edge states.   The most relevant   gauge invariant operators that can be generated on integrating out the massive fermions are CS operators involving the two gauge fields.  In particular we obtain the following contribution to the effective Lagrangian\footnote{Our metric convention is $\eta^{\alpha\beta}=\text{diag}\{1,-1,-1\}$ with $\epsilon_{012} = 1$ and Dirac matrices satisfying $\{\gamma^\alpha,\gamma^\beta\} = 2\eta^{\alpha\beta}$.}, 
  \beq
 \CL_\text{CS} &=& \frac{   \epsilon^{\alpha\beta\gamma}}{4\pi}  \Bigl[n_\psi A_\alpha \partial_\beta  A_\gamma 
 + n_\chi Z_\alpha \partial_\beta  Z_\gamma  
 \cr &&\qquad\quad
  + n_\omega (A_\alpha + Z_\alpha)\partial_\beta  (A_\gamma+ Z_\gamma)\Bigr]+\ldots\ ,
 \eqn{CS}\eeq
 where the ellipses refers to higher derivative operators, including the Maxwell terms.
 These $P$ and  $T$ violating CS operators are proportional to the signs of the fermion and Pauli-Villars masses, which are odd under $P$ and $T$; the opposite sign mass for the Pauli-Villars field, in conjunction with its opposite statistics, ensures that its contribution adds to the rather than cancels for each fermion.   At this point it should be no surprise then that the theory will exhibit the FQHE, since  \eq{CS} is an example of the effective description for the FQHE in condensed matter systems  \cite{wen1992classification}.

With the Chern-Simons operators being linear in derivatives and the Maxwell terms quadratic, the gauge boson propagators develop  gauge-invariant poles at nonzero mass \cite{Deser:1981wh}. With $g^2\gg e^2$,  the $Z$ boson is heaviest with mass 
\beq
M_Z = g^2(n_\chi+n_\omega)/(2\pi)+O(e^2/g^2)\ ,
\eeq
 and so we integrate it out of the theory to create the effective theory for the photon.  Since we are looking for a theory with momenta $k$ satisfying $M_Z \gg k > M_\gamma$, we can take $g\to\infty$ at this point in our calculations, ignoring the irrelevant Maxwell term for the $Z$.  Since the theory is gauge invariant, we need to introduce a gauge fixing term, however, $\frac{1}{2\xi}( Z^\alpha\partial_\alpha\partial_\beta Z^\beta)$.  The $Z$-dependent part of the Lagrangian may then be written as
 \beq
\CL_Z &=& \half\Bigl[\left(Z- \frac{n_\omega}{2\pi}A \Delta \CQ^{-1}\right)  \CQ  \left(Z- \frac{n_\omega}{2\pi} \CQ^{-1} \Delta A\right)  
\cr
&&\cr &&
\quad
-\left( \frac{n_\omega}{2\pi}\right)^2 A \Delta \CQ^{-1} \Delta A\Bigr]\ ,
\eqn{eft1}\eeq
with definitions
\beq
\CQ &=& -\frac{(n_\chi+n_\omega)}{2\pi}\Delta + \frac{\Xi }{\xi } \ ,\cr
\Delta^{\alpha\beta} &=& \epsilon^{\alpha\beta\gamma}\partial_\gamma \ ,\qquad
\Xi^{\alpha\beta}= \partial^\alpha\partial^\beta .
\eeq
The propagator is then computed to be 
\beq
\CQ^{-1} = \frac{2\pi}{(n_\chi+n_\omega)}    \frac{\Delta}{\partial^2}+\frac{\xi \,\Xi}{(\partial^2)^2}\ ,
\eeq
from which follows
$
\Delta \CQ^{-1} \Delta  =-   2\pi\Delta /(n_\chi+n_\omega) 
$.
Therefore, after integrating out the $Z$, we are   left with the low energy effective theory for the photon,
\beq
\CL_\text{EFT}
&=& \frac{1}{4e^2} F_{\alpha\beta} F^{\alpha\beta} +\nu\, \frac{   \epsilon^{\alpha\beta\gamma}}{4\pi} A_\alpha \partial_\beta  A_\gamma +\ldots\ ,\cr  &&\cr
\nu&=&\left(n_\psi+  \frac{n_\chi n_\omega }{n_\chi + n_\omega}\right)\ ,
\eqn{EFT}
\eeq
where  the ellipses represent irrelevant higher derivative operators.  The Hall current for this system with conventional normalization of the gauge field is then given by 
\beq
J^\alpha= \nu\frac{ e^2 }{4\pi} \epsilon^{\alpha\beta\gamma}  F_{\beta \gamma}
\eqn{JHall}
\eeq
implying a fractional Hall conductivity 
$
\sigma_{xy} = \nu e^2/h$.
In general our expression for $\nu$ takes non-integer values, as displayed for several examples in Table~\ref{tab:table2}.

\begin{table*}
\caption{\label{tab:table2} Properties of the low energy spectrum for various values of the UV flavor numbers $\{n_\psi,\,n_\chi,\,n_\omega\}$.  The quantity $\nu$ from \eq{EFT} is the analogue of the filling fraction in condensed matter systems while $\{q_\psi',\,q_\chi',\,q_\omega'\}$ are the charges of the massless chiral edge states;  $q_\phi$ and $\alpha_\phi$ are the charge and statistics of bulk excitations arising from  a boson field $\phi$ in the UV with $U(1)\times U(1)$ charges $q_Z=1$, $q_A=0$. }
\begin{ruledtabular}
\begin{tabular}{ccccc}
$\{n_\psi,\,n_\chi,\,n_\omega\}$ & $\nu$ &$\{q_\psi',\,q_\chi',\,q_\omega'\}$ & $q'_\phi$    & $\alpha_\phi$ \\ \hline
$\{1,-1,-2\}$ & $1/3$ & $\{1,\,-2/3,\,1/3\}$ & $-2/3$ & $4/3$ \\
$\{0,1,1\}$ & $1/2$ & $\{1,\,-1/2,\,1/2\}$&$-1/2$& $1/2$\\
$\{0,1,2\}$ & $2/3$ & $\{1,\,-2/3,\,1/3\}$ &$-2/3$ & $2/3$\\
$\{2,1,1\}$& $5/2$ &$\{1,\,-1/2,\,1/2\}$&$-1/2$ &$1/10$\\
\end{tabular}
\end{ruledtabular}
\end{table*}

 \section{Fractionally charged chiral surface modes and bulk excitations}
 
When the system has a boundary, gauge invariance requires that there must be fractionally charged chiral edge modes whose $U(1)$ anomaly cancels the divergence of the  Chern-Simons current in \eq{JHall}.  Indeed, if one turns off the gauge interactions, one can show that a chiral edge state exists for each flavor of fermion with chirality proportional to the sign of the bulk fermion mass  \cite{Jackiw:1975fn}, but from their charge assignments in Table~\ref{tab:charges} it appears that they can only carry  integer charge under the $U(1)_A$ gauge group. To resolve this puzzle we consider the theory on the half-space $x^2\ge 0$ with boundary condition for both the fermions and Pauli-Villars fields of type $i$
\beq
P_i^+\Psi_i\bigl\vert_{x^2=0} = P_i^-\Psi_i\bigl\vert_{x^2=\infty}=0\ ,\ \  
P_i^\pm =\frac{\left(1 \pm \epsilon(n_i)\Gamma\right)}{2}  ,
\eqn{BC}\eeq
where $n_i$ is the flavor number of that fermion (recall that for each fermion the sign of $n$ reflects the underlying sign chosen for its mass), and $\Gamma = i\gamma^2$ is the hermitian chiral matrix with eigenvalues $\pm 1$.  Massless chiral edge state correspond to solutions   to $(i\gamma^2\partial_2 - m_i)\Psi_i = 0$ consistent with the boundary condition \eq{BC}. Such  states exist for each fermion with chirality given by $-\epsilon(n_i)$. The Pauli-Villars fields of type $i$ have opposite sign mass (but the same $n_i$)  and the boundary condition forbids a zeromode solution for them.
 
    To compute the effective theory   for the half-space we must include the massless edge modes,  adding to the UV action
$\int d^3x (A_\mu \CJ^\mu_A + Z_\mu \CJ^\mu_Z)$, where the currents have the form dictated by the charges  $q_{A,Z}$ given in Table~\ref{tab:charges}
\beq
\CJ_{A}^\mu &=& \left(\CJ_\psi^\mu+\CJ_\omega^\mu \right)\ ,\quad
\CJ_{Z}^\mu = \left(\CJ_\chi^\mu+\CJ_\omega^\mu \right)\ ,\cr
\CJ_i^\mu&=&\delta(x^2) \bar \Psi_i \gamma^\mu  P^-_i \Psi_i\
 ,\quad i=\psi,\chi,\omega\ ,
\eeq
where   $\mu=\{0,1\}$ designates the $d=1+1$ coordinates on the mass defect, and the $\Psi_i$ only include fermions, not Pauli-Villars fields. 
Integrating out the $Z$ boson therefore induces couplings between $\CJ_{Z}^\mu$ and the photon, and we must add to the effective theory in \eq{EFT} the edge mode interactions
\beq
\CL^\text{zm}_\text{EFT} &=&   \frac{n_\omega}{4\pi} \left(A^\alpha \left(\Delta \CQ^{-1}\right)_{\alpha\beta} \CJ^\beta_Z +\CJ_Z^\alpha \left( \CQ^{-1}\Delta\right)_{\alpha\beta}  A^\beta\right) \cr &&
+A_\mu  \CJ^\mu_A\ ,
\eqn{Jlag}\eeq
 with terms quadratic in  $\CJ_Z$ vanishing. If we combine the  contributions to the photon current from the first two terms above with the bulk contribution  from the last term in \eq{eft1}, we find that the photon current induced by integrating out the $Z$ is given by
\beq
J^\alpha_\text{induced} =     \frac{n_\omega}{2\pi} \,\left( \Delta \CQ^{-1}\left[\CJ_Z - \frac{n_\omega}{2\pi} \Delta A\right]\right)^\alpha\ .
\eeq
The quantity in parentheses is the conserved current identified by Callan and Harvey \cite{Callan:1984sa}, the divergence of the bulk current at the boundary being cancelled by the anomaly of the edge state current \footnote{ The cancellation requires one use the covariant anomaly for the edge states   instead of the consistent anomaly, which is half as large; the discrepancy is resolved by the generation of additional gauge field terms  at the surface  \cite{Naculich:1987ci}.}. The operator acting on this current is 
\beq
(\Delta \CQ^{-1})^{\alpha\beta}   =-\frac{2\pi}{n_\chi+n_\omega}\left(\eta^{\alpha\beta}-\frac{\partial^\alpha\partial^\beta}{\partial^2}\right)\ , 
\eeq
which is recognized as being proportional to the projection operator onto physical states. Without altering the physics we can make the interaction local by adding  a similar term involving the projection operator onto unphysical longitudinal states,  since the added interaction is proportional to the divergence of the conserved current, allowing us to replace
\beq
(\Delta \CQ^{-1})^{\alpha\beta} \to -\frac{2\pi}{n_\chi+n_\omega} \eta^{\alpha\beta}\ . 
\eeq
 Adding the resultant induced current to the direct zeromode contribution $ \CJ^\mu_A$ in \eq{Jlag} yields the total interaction of the photon with the edge states
\beq
\CJ_{A,\text{EFT}}^\alpha =  \left(q_\psi'\CJ_\psi^\mu+ q_\chi'  \CJ_\chi^\mu +q_\omega' \CJ_\omega^\mu  \right) 
\eqn{jedge}\eeq
with
\beq
q_\psi' = 1,\ \  q_\chi' = -\frac{  n_\omega}{n_\chi+n_\omega},\ \   q_\omega' = 1  -\frac{  n_\omega}{n_\chi+n_\omega}\ ,
\eqn{fracq}\eeq
taking into account the $(n_\omega/4\pi) $ prefactor in \eq{Jlag}. We see therefore that in the low energy theory, the chiral $\chi$ and $\omega$ edge states carry fractional charge.  For example, 
our $\nu=1/3$ example from Table~\ref{tab:table2} 
yields the fractional charges
$q_\chi' = -2/3$ and 
$q_\omega' = 1/3$ for the chiral $\chi$ and $\omega$ edge states respectively.

For our effective theory to be gauge invariant,  the $d=1+1$ anomaly at the edge of the sample with these fractionally charged edge modes must correctly cancel the boundary divergence of the Hall current \eq{JHall}. 
 Indeed this is the case since
 \beq
 \sum_{i=\psi,\chi,\omega} n_i\left(q'_i\right)^2  = \left(n_\psi+  \frac{n_\chi n_\omega }{n_\chi + n_\omega}\right)=\nu\ .
  \eeq

Bulk  excitations in the IR can be studied by adding additional light fields to the theory. Consider, for example,
a bosonic field $\phi$ of mass $m_{\phi}$ with $m_{\phi}< M_Z$ and  charges  $q_A=0$, $q_Z=1$. After integrating   out the $Z$ field, 
 $\phi$ couples to the photon with induced fractional charge given by 
\beq
q_{\phi}'=q_\chi' = -\frac{n_{\omega}}{n_{\chi}+n_{\omega}}.
\eeq
A magnetic flux  $2\pi q'_\phi/\nu$ attaches to the particle in the IR, and so in the usual way the resulting Aharanov-Bohm phase modifies the boson statistics so that the 2-boson state satisfies $\ket{\psi_1\psi_2} = e^{i\alpha_\phi}\ket{\psi_2\psi_1} $ with
\beq
\alpha_\phi =\frac{ (q_\phi')^2}{\nu} = 
\frac{n_{\omega}^2}{(n_{\chi}+n_{\omega})(n_{\chi}n_{\psi}+n_{\omega}n_{\psi}+n_{\chi}n_{\omega})}.
\eeq
Examples of the fractional coupling and statistics for these excitations in Table~\ref{tab:table2}. Different initial $q_{A,Z}$ charges will lead to  bulk excitations with different behavior in the IR theory. 

 \section{Fractional Quantum Spin Hall Effect}
 The quantum spin Hall effect (QSHE) occurs when the current in the bulk transports global quantum numbers but not gauged charges, in contrast with the classical Hall current.  In the language of quantum field theory, this means that the edge states form an anomaly-free representation of the gauge group, while various flavor symmetries can have gauge anomalies.
This effect  was analyzed in Ref.~\cite{Kaplan:1992bt}, where it was suggested that by physically separating topologically protected edge states in  gauge anomaly free representations from their conjugate counterparts at the opposite edge, one could turn the problem  of finding a  nonperturbative regulator  for chiral gauge theories into finding the appropriate phase of a lattice theory with spatially dependent interactions\footnote{This idea is still being pursued in various forms \cite{Wen:2013ppa,Wang:2013yta,DeMarco:2017gcb,Wang:2018ugf,grabowska2016nonperturbative,grabowska2016chiral}.}.  By construction, such materials only have CS (Hall) currents which carry flavor quantum numbers in the bulk. The simplest example is a material with a massless Dirac fermion at each edge, such as seen in condensed matter systems \cite{hasan2010colloquium}, but less trivial examples of anomaly cancellation have been investigated numerically, such as the 3-4-5 model \cite{Jansen:1992yj} which has a chiral representation on the boundary.

It is evident that a fractional quantum spin Hall effect is possible,  where the Hall currents only transport fractional flavor charges.  In order to not have transport of electric charge we must have the coefficient $\nu$ of the Hall current in \eq{JHall} vanish, namely
\beq
 \nu = \left(n_\psi+  \frac{n_\chi n_\omega }{n_\chi + n_\omega}\right) = 0\ .
\eqn{anomcancel} \eeq
 We can then consider the transport of any linear combination of the many  U(1) flavor currents in the model, normalized so that the fermions in the UV theory carry integer charge,  and compute its Hall current.  As a specific example, consider the flavor current $j^\alpha$ of a single one of the $\vert n_\chi\vert$ $\chi$ fermions, e.g. $\chi_1$.  We can introduce a source $f_\alpha$  in the UV theory coupled to the current $\bar\chi_1\gamma^\mu\chi_1$ and compute the effective theory with all fields integrated out except for $f_\alpha$, $A_\alpha$ and the chiral edge states. The flavor current in the IR can be obtained by differentiating this action with respect to    $f_\alpha$, then setting $f_\alpha=0$.  A nontrivial result for the Hall conductivity of our flavor current will be obtained when the effective theory has an induced Chern-Simons coupling between $f_\alpha$ and $A_\alpha$. This is the case with our particular example, and we find that the Hall conductivity for this flavor is
 \beq
\sigma^{(\chi_1)}_{xy} =-2\frac{n_\omega}{n_\omega+n_\chi} \, \frac{e^2}{h} = 2\frac{n_\psi}{n_\chi} \, \frac{e^2}{h}\ .
\eqn{FQSHE}
\eeq
For example, if we set $n_\psi =2$,  $n_\chi = -6$ and $n_\omega=-3$ we solve the anomaly cancellation equation \eq{anomcancel}, while the Hall conductivity for $\chi_1$ number given by \eq{FQSHE} is  $\sigma^{(\chi_1)}_{xy} =- \frac{2}{3} \frac{e^2}{h}$, and this fractional Hall conductivity will be accompanied by massless edge modes carrying fractional $\chi_1$-flavor number as well, to ensure flavor current conservation\footnote{FQSHE has been previously discussed for theories possessing time-reversal symmetry in Ref.~\cite{levin2012classification}}.

 \section{Topological phases}
 
  Ref.~\cite{Golterman:1992ub} explained the quantization of the CS coefficient as being topological in origin, equal to the winding number of the map provided by the fermion dispersion relation from momentum space to the Dirac sphere.  This winding number only has meaning in the IR and depends on the fermions being massive (gapped) in the bulk.  Topological phase transitions can occur at points or surfaces in parameter space where the theory becomes gapless.  It was the observation of discontinuous jumps in the spectrum of chiral edge states in lattice quantum field theory at particular parameter values  that  motivated that work.  
 
 In the example we have analyzed here, each fermion of mass $M$ and its accompanying Pauli-Villars regulator of mass $\Lambda$ combined contribute a   factor to the coefficient of the CS operator proportional to $(M/|M| - \Lambda/|\Lambda|) = \pm 2$, given our choice that $\Lambda/|\Lambda| = -M/|M|$.  Evidently if we had chosen the Pauli-Villars mass to have the same sign as the fermion, the two contributions would cancel, the magnitude of the fractional Hall current would change, and the charges of the chiral edge modes would be different. Therefore a topological phase transition can be seen by varying the fermion mass continuously from $M \to -M$ where the transition occurs at $M=0$ and the  CS operator coefficient changes by $\pm2$.    However, our  boundary condition \eq{BC} would also change discontinuously which makes this example difficult to analyze.  So instead of a half space,   consider the full space, but where the fermion mass $M$ to be positive for one sign of  $x^2$  and negative for the other, so that the chiral edge modes live on the ``domain wall" at $x^2=0$.  The Pauli-Villars mass is taken to be constant everywhere so that there are no massless edge states with negative norm.  Now we see that the sum of fermion and Pauli-Villars contributions add on one side of the circle and cancel on the other, so that the transformation $M \to -M$ has the effect of  a parity transformation, moving the nonzero CS current from one side of $x^2=0$ to the other and flipping the chirality of the edge-modes, but not rendering the topology trivial.

Instead of Pauli-Villars regularization we could employ a lattice regularization with Wilson fermions, as analyzed explicitly in Ref.~\cite{Golterman:1992ub}. In this case one finds a much richer topological phase diagram, where one can vary the ratio of the fermion mass to Wilson coefficient $M/r$ for each fermion \footnote{The  Wilson parameter $r$ refers to  the coefficient of the $\bar\psi D^2 \psi$ operator, where $D^2$ is the gauge covariant lattice Laplacian.}, toggling the number and chirality of edge states for each flavor through the Pascal Triangle numbers $0,1,2,1,0$ with alternating chirality; see for example \cite{Kaplan:2009yg} for a discussion and  phase diagram.   Since this can be done individually for each fermion, we see that the theory above has a very large number of possible topological phases when formulated in terms of Wilson fermions on a lattice.  There have been  several recent papers investigating the topological phase diagram of $U(1)$ gauge theories on the lattice in $2+1$ dimensions  \cite{Magnifico:2018wek,Bermudez:2018eyh,Tirrito:2018bui,Magnifico:2019ulp},
and  many of those results can be extended to the case we consider here with a $U(1)\times U(1)$ gauge group.

 \section{Discussion}

By building on existing ideas for effective field theory descriptions of the FQHE in condensed matter systems we have explicitly constructed relativistic UV completions that exhibit the same phenomenon of charge fractionalization. These theories have only integer charged fermions and gauge fields in the UV, and  they exhibit a topological phase structure which (as should be expected) is dependent on the specifics of the UV regularization used.  The advantage of considering such theories is  the simplicity and perturbative nature of the UV physics that allows one to easily see how the IR physics arises. There are many ways one can extend this approach, such as by extending the gauge group to include more abelian or non-abelian gauge fields, to incorporate spontaneous symmetry breaking, and by working in higher dimension.
It is hoped that in pursuing this program it may be possible to further our understanding of topological phenomena and how they can arise.

\acknowledgements{
	This research was supported in part  by DOE grant No.~DE-FG02-00ER41132 and Iowa State University startup funds. We thank Andreas Karch for useful conversations.
	}

\bibliography{FQHE}
\end{document}